\documentstyle[12pt]{article}
\newcommand{\II}{{\cal I}}

\newcommand{\NN}{{\cal N}}

\newcommand{\wt}{\widetilde}
\newcommand{\wb}{\bar}

\newcommand{\vr}{\vec r}

\newcommand{\be}{\begin{equation}}
\newcommand{\ee}{\end{equation}}
\newcommand{\ben}{\begin{eqnarray}\displaystyle}
\newcommand{\een}{\end{eqnarray}}
\newcommand{\refb}[1]{(\ref{#1})}
\newcommand{\p}{\partial}

\begin{document}

{}~ \hfill\vbox{\hbox{hep-th/9708002}\hbox{MRI-PHY/P970720}
\hbox{Revised}}\break

\vskip 3.5cm

\centerline{\large \bf Strong Coupling Dynamics of Branes}

\centerline{\large \bf from M-theory}

\vspace*{6.0ex}

\centerline{\large \rm Ashoke Sen\footnote{On leave of absence from 
Tata Institute of Fundamental Research, Homi Bhabha Road, 
Bombay 400005, INDIA}
\footnote{E-mail: sen@mri.ernet.in, sen@theory.tifr.res.in}}

\vspace*{1.5ex}

\centerline{\large \it Mehta Research Institute of}
 \centerline{\large \it   Mathematics and Mathematical Physics}

\centerline{\large \it Chhatnag Road, Jhusi, Allahabad 221506, INDIA}

\vspace*{4.5ex}

\centerline {\bf Abstract}

We study some aspects of the strong coupling dynamics of
Dirichlet six branes, anti- six branes, and orientifold planes by
using the equivalence of type IIA string theory and $M$-theory on
$S^1$. In the strong coupling limit there
exists static configuration of brane and anti-brane at arbitrary
separation, suspended in an external magnetic field. The mass of 
the open string stretched between the
brane and the anti-brane approaches a finite positive value even when
the branes coincide. Similar result is obtained for a Dirichlet six brane
on top of an orientifold six plane.  We also derive the anomalous 
gravitational interaction on the brane and the orientifold plane from
$M$-theory.

\vfill \eject

\baselineskip=18pt

\section{Introduction and Summary}

Many known properties of Dirichlet six branes in type IIA
string theory\cite{GIMPOL,HORGR,POLC} have been 
reproduced\cite{RUBACK,KK,IMA,ENH} by using 
the known 
identification\cite{DUK}-\cite{GREH}
of these 
branes with Kaluza-Klein monopoles\cite{SOR,GP} in 
$M$-theory. In this paper we shall use this identification to
derive some properties of these D-branes
which cannot be derived using the standard string 
perturbation theory.

The first system we shall analyse is the brane $-$ anti-brane
configuration. This corresponds to a complicated interacting 
system in perturbative string theory due to the appearance of
tachyonic open string states for sufficiently small separation
between the brane and the anti-brane\cite{BANSUS,GRGUP,PERI}. 
However, by mapping it to the known Kaluza-Klein dipole
solution\cite{GP,GIBHAW} in $M$-theory, we
show that the dynamics simplifies in the strong coupling limit of
string theory.  The dipole
solution in eleven dimensional supergravity theory can be
interpreted as a static brane $-$ anti-brane configuration in type
IIA string theory, suspended in an external magnetic field\cite{DOW}.
We calculate the mass of the open string stretched between
the brane and the anti-brane by identifying it with the $M$-theory
membrane stretched along the bolt of the Kaluza-Klein dipole
solution. While this answer agrees with
the expected answer from string theory for large separation
between the branes, it  approaches a finite value proportional
to $m_{string}g_{string}$ when the separation between the brane and
the anti-brane vanishes. In perturbative string theory, 
the classical contribution to the mass vanishes in this limit,
and quantum fluctuations on the open string makes this into
a tachyonic state\cite{BANSUS}.

The second system that we shall analyze will be a D- six brane in
the presence of an orientifold six plane\cite{GIMPOL,SAGN}. 
For this we use the
identification of the orientifold plane with the Atiyah-Hitchin
space of $M$-theory\cite{SEIWITTH,ENH}. In string perturbation
theory, the classical mass of the open string, stretched between
the D- six brane and its image, is proportional to the distance
between the D-brane and the orientifold plane. When the D-brane
is on top of the orientifold plane, the classical mass vanishes,
but the massless states get projected out, and the lowest mass
state from this sector has mass of order unity in the string
scale. From the $M$-theory viewpoint, on the other hand, this
configuration of coincident
D-brane and orientifold plane is described by
the double cover of the Atiyah-Hitchin space\cite{SEIWITTH}, 
and the open
string stretched between the D-brane and its image is represented
by a membrane wrapped around the bolt of the Atiyah-Hitchin
space. The classical mass of this state is proportional to the
area of the bolt, and when converted to the string scale, this
mass again turns out to be proportional to $m_{string}g_{string}$.
This is an exact result in the strong coupling
limit of string theory.

In the final section, we shall deviate somewhat from the main
theme of the paper, and use the $M$-theory $-$ IIA correspondence 
to derive a known result, $-$ the anomalous gravitational
coupling of the six brane. This
result was derived in refs.\cite{BSV,IBR} by other methods. 
We derive this by starting from the $C\wedge X_8$ coupling in
$M$-theory\cite{VWONE,VAFWIT}, and integrating it 
over the transverse
space of the six brane in $M$-theory. The same method can be used
to prove the existence of a similar coupling on the orientifold
plane. This has been derived using different method in
a recent paper by Dasgupta, Jatkar and Mukhi\cite{DJM}.

\section{Brane $-$ Anti-brane Configuration}

Our starting point for studying the brane $-$ anti-brane 
configuration will be the Kaluza-Klein dipole solution given in
\cite{GP}. The solution embedded in eleven dimensions
is described by the metric:
\ben \label{e1}
ds^2 &=& -dt^2 + \sum_{m=5}^{10} dy^m dy^m + (r^2 -a^2
\cos^2\theta) [\Delta^{-1} dr^2 + d\theta^2] \nonumber \\
&& + (r^2 - a^2 \cos^2\theta)^{-1} [ \Delta (dx^4 + a\sin^2\theta
d\psi)^2 + \sin^2\theta \big( (r^2-a^2)d\psi - a dx^4\big)^2]\, ,
\een
where,
\be \label{e2}
\Delta = r^2 - 2Mr - a^2\, .
\ee
$M$ and $a$ are parameters labelling the solution. 
Let us define,
\be \label{e3}
\kappa = {\sqrt{M^2 + a^2}\over 2M(M+\sqrt{M^2+a^2})}\, ,
\qquad \Omega = \kappa {a \over \sqrt{M^2 + a^2}}\, ,
\ee
and
\be \label{e4}
\phi = \psi -\Omega x^4\, .
\ee
The solution \refb{e1}
is regular if $r\ge r_0$ where,
\be \label{e5}
r_0 = M + \sqrt{M^2 + a^2}\, ,
\ee
and $x^4$ and $\phi$ have periodicities $2\pi/\kappa$ and $2\pi$
respectively. If we use $x^4$ and $\phi$ as independent angular
coordinates, then the surface $r=r_0$ is a fixed point of the killing
vector $\p/\p x^4$ and represents a bolt.
\refb{e1} represents a valid solution of $M$-theory in the limit
of large $M$, when the curvature associated with the solution 
is small, and hence the higher derivative terms in the Lagrangian
are not important.

In appropriate coordinate system, this describes a magnetic
monopole $-$ anti-monopole pair suspended in an external magnetic
field\cite{DOW}. In order to see this, we shall first analyze
this solution for $a>>M$. 
First of
all, note that in the region $(r-a)>>M$, $\Delta$  can be
approximated as:
\be \label{emod1}
\Delta \simeq (r^2-a^2)\, .
\ee
Substituting this into the metric \refb{e1}, and defining new
coordinates $\rho$ and $\bar\theta$ through the relations:
\be \label{emod2}
\rho \sin\bar\theta = \sqrt{r^2-a^2} \sin\theta\, ,
\qquad \rho\cos\bar\theta = r \cos\theta\, ,
\ee
we can bring the metric \refb{e1} into the form
\be \label{emod3}
ds^2 = -dt^2 + \sum_{m=5}^{10} dy^m dy^m + (dx^4)^2 
+ d\rho^2 + \rho^2 (d\bar\theta^2 + \sin^2\bar\theta d\psi^2)\, .
\ee
This appears to be the standard flat metric on $R^{9,1}\times S^1$.
However the background is
non-trivial because of the twisted boundary condition on $x^4$
and $\psi$ (as $x^4$ gets translated by $2\pi/\kappa$, $\psi$ 
must get translated by $2\pi\Omega/\kappa$ in order to have an
identification of points in space-time). There are many ways of
choosing independent angular coordinates $-$ one of them being
$x^4$ and $\phi$ defined in eq.\refb{e4} $-$ but we shall take the
independent coordinates to be $x^4$ and 
\be \label{emod4}
\wt\phi = \phi + \kappa x^4=\psi + (\kappa -\Omega) x^4\, .
\ee
The asymptotic space-time may then be interpreted as $M$-theory
compactified on $S^1$ labelled by $x^4$, in the presence of a
magnetic flux\cite{GIBT,GIBM,GIBO}. The magnetic field on the $z$
axis is given by:
\be \label{emod5}
B = \Omega - \kappa \simeq -{M\over 4a^2}
\quad \hbox{for} \quad a>>M\, .
\ee

We shall now turn to the interior region where $(r-a)$ is of
order $M$. In order to properly interpret
the solution, we also consider the single monopole solution\cite{SOR,GP}:
\be \label{e7}
ds^2 = -dt^2 + \sum_{m=5}^{10} dy^m dy^m + ds_{TN}^2\, ,
\ee
\be \label{e7a}
ds_{TN}^2=
(1+{4m\over r}) d\vr^2 +
(1+{4m\over r})^{-1} (dx^4 + 4m (1-\cos\theta) d\phi)^2 \, ,
\ee
where $x^4$ has periodicity $16\pi m$. This represents a 
Kaluza-Klein monopole of magnetic charge $4m$. We can 
interpret both the monopole and dipole solution as solutions in
the same theory by matching the $x^4$ periodicity in the two
theories.  This gives,
\be \label{e8}
m = {1\over 8\kappa}\, .
\ee
We shall now show that the solution \refb{e1} 
corresponds to a monopole-antimonopole pair, separated by a
distance $2a$ for large $a$. This will be done by
showing that i) in the limit of large $a$, the metric around $r=r_0$,
$\theta=0$ ($\theta=\pi$) reduces to that of a Kaluza-Klein 
anti-monopole
(monopole), and ii) for finite $a$, but close to the point
$r=r_0$, $\theta=0$ ($\theta=\pi$) the metric agrees with the
metric
close to a Kaluza-Klein anti-monopole (monopole).
For this we introduce a new set of coordinates:
\be \label{epp}
(r_0-M)\sin^2\theta = \wt\rho (1-\cos\wt\theta), \qquad 
2(r-r_0) = \wt\rho(1+\cos\wt\theta)\,  .
\ee
First we focus on the region near $(r=r_0, \theta=0)$
and take the $a\to\infty$ limit keeping $(r-r_0, \theta\sqrt a)$
finite. 
In this limit, the metric \refb{e1} reduces to:
\ben \label{epp2}
ds^2 &=& -dt^2 +\sum_{m=5}^{10} dy^m dy^m + \Big(1 +
{M\over\wt\rho}\Big)^{-1} \big(dx^4 - M(1-\cos\wt\theta)
d\wt\phi\big)^2\nonumber \\
&& + \Big(1+{M\over \wt\rho}\Big) (d\wt\rho^2 + \wt\rho^2
d\wt\theta^2 + \wt\rho^2\sin^2\wt\theta d\wt\phi^2)\, .
\een
The coordinate ranges are $0\le\wt\rho<\infty$,
$0\le\wt\theta\le\pi$. Furthermore, 
$x^4$ and $\wt\phi$ can be taken to be independent angular
coordinates with periods $4\pi M$ and $2\pi$ respectively. Thus
we see that \refb{epp2} represents the metric of a Kaluza-Klein
anti-monopole with magnetic charge $-M$. Similar analysis can be
carried out near $(r=r_0, \theta=\pi)$.

On the other hand if we keep $a$ finite 
and examine the solution
close to the region $\wt\rho=0$, the metric reduces to:
\ben \label{epp3}
ds^2 &=& -dt^2 +\sum_{m=5}^{10} dy^m dy^m + 
(2\kappa \wt\rho) \big(dx^4 - (2\kappa)^{-1}(1-\cos\wt\theta)
d\wt\phi\big)^2\nonumber \\
&& + (2\kappa\wt\rho)^{-1} (d\wt\rho^2 + \wt\rho^2
d\wt\theta^2 + \wt\rho^2\sin^2\wt\theta d\wt\phi^2)\, .
\een
Using eq.\refb{e8}, we can recognise it as the 
metric given in eq.\refb{e7}, \refb{e7a} 
close to the Kaluza-Klein anti-monopole. This is a reflection of
the fact that in the choice of coordinates we have made, the
metric near $r=r_0$, $\theta=0$ represents an anti- self-dual
NUT\cite{GIBHAW}. Similarly, the metric near $r=r_0$,
$\theta=\pi$ represents a self-dual NUT and hence a magnetic
monopole.

As has already been stated, the Kaluza-Klein monopoles in
$M$-theory can be interpreted as D- six branes of type IIA 
string theory\cite{TOWNS}. Thus the solution \refb{e1} describes
a static D6-brane $-$ anti- D6-brane configuration of type IIA
string theory.
The distance
between the brane and the anti-brane can be defined to be the
geodesic distance between the points $(r_0,\theta=0)$ and
$(r_0,\theta=\pi)$. This is given by
\be \label{eqq2}
l=\int_0^\pi \sqrt{r_0^2-a^2\cos^2\theta} \, d\theta \simeq 2a
\qquad \hbox{for} \quad a>>M\, .
\ee
Existence of brane anti-brane solution at arbitrary separation
$l$ (parametrized by $a$) shows that the static force between the
brane and the anti-brane vanishes. Since the gravitational and
electromagnetic interaction between a brane and an anti-brane are
both attractive\cite{POLC,BANSUS,GRGUP,PERI}, there is a net
attractive force between them. This is cancelled by the
repulsive force between the
brane and the antibrane, induced by the external magnetic field.
For monopole and anti-monopole carrying magnetic charges $\pm M$
and separated by a distance $2a$, this repulsive force is
equal to $-2MB\simeq M^2/2a^2$ for large $a$. On the other hand, the
magnetic 
attraction betwen this pair is $M^2/4a^2$. The gravitational (and
scalar induced) attraction is equal in magnitude to the magnetic
attraction. Thus we see that the net attractive force cancels the
net repulsive force, giving rise to a static configuration.

We shall now identify the open string state stretched between the
brane and the anti-brane, and calculate its classical mass. In
analogy with the results of ref.\cite{ENH} this state must be
represented by an $M$-theory membrane wrapped on a suitable two
cycle. For the dipole solution, such a two cycle is provided by
the surface $r=r_0$, which appears as a bolt if we choose $\phi$ and
$x^4$ as the independent angular coordinates. 
$\theta$ and $\phi$ are good coordinates on the
bolt, and the metric on the bolt is given by:
\be \label{e10}
ds_B^2 = (r_0^2 - a^2 \cos^2\theta)^{-1} (r_0^2-a^2)^2
\sin^2\theta d\phi^2 + (r_0^2 - a^2\cos^2\theta) d\theta^2 \, .
\ee
The area of this surface in this metric is given by:
\be \label{e11}
A = 4\pi (r_0^2-a^2)= 8\pi M (M + \sqrt{M^2+a^2})\, .
\ee
For large $a$, this reduces to $8\pi Ma$, which is simply the
product of the period $4\pi M$ of $x^4$, and the separation $2a$
between the brane $-$ anti-brane pair. If $T_M$ denotes the
membrane tension in $M$ theory then this state has mass $T_M A$.
Since $T_M$ times the period of the $x^4$ direction can be
identified as the string tension of type IIA theory, we see that
for large separation between the branes, the mass of the string
is given by the product of the string tension and the separation,
$-$ as expected from perturbative string analysis.

In fact, not only is the mass formula reproduced correctly for
large $a$, but the shape of the membrane also has the right 
form so as to be interpreted as 
a string of length $2a$. To see this, let us note from \refb{e10}
that for fixed $\theta$, the radius of the $\phi$ direction is
given by:
\be \label{e12}
{(r_0^2-a^2)\sin\theta\over
\sqrt{r_0^2-a^2\cos^2\theta}}\int_0^{2\pi} d \phi 
\simeq  4\pi M \qquad \hbox{for large $a$, \quad $\theta\ne 0, \pi$}\, .
\ee
Thus away from the north and the south pole ($\theta=0,\pi$)
the bolt has the shape of a cylinder of radius $4\pi M$. As seen
from \refb{e1}, the compact
direction on the cylinder is spanned by $\phi=\wt\phi+\kappa x^4$ 
at fixed value of $x^4+a\sin^2\theta\psi$. For large $a$ and finite
$\theta$, this means that the compact direction is
spanned by $x^4$ at fixed $\wt\phi$. The
length of this cylinder, stretched along the $\theta$ direction,
is equal to $l$ defined in \refb{eqq2}.  A
membrane wrapped around such a surface clearly looks like an
elementary type IIA string stretched from $\theta=0$ to
$\theta=\pi$, covering a distance $2a$. 

When $a$ is of the order of $M$, the ten dimensional
interpretation is not reliable, reflecting the fact that the magnetic
field required to hold the brane $-$ anti-brane pair apart
produces curvature in the non-compact direction which is of the
same order as the size of the compact
direction\cite{DOW,GIBT,GIBM,GIBO}. Note, however, that the solution 
is completely non-singular in eleven dimensions.
For $a=0$, we again have
a ten dimensional interpretation of the solution if we take
$\phi$ instead of $\wt\phi$, and $x^4$ as the independent angular
coordinates. In this case the asymptotic space has the structure
of $R^{9,1}\times S^1$ with flat metric. Furthermore the metric
does not have any $g_{4\phi}$ component, showing that the
solution has zero magnetic field. Thus it is natural to
interpret this configuration as a coincident brane-anti-brane
pair. The alert reader may raise several objections to this
proposal. First of all, note that if we continue to use $\wt\phi$
and not $\phi$ as the azimuthal coordinate of the transverse
space, then the solution for $a=0$ can be interpreted as a
monopole $-$ anti-monopole pair at the two poles of the surface
$r=r_0$\cite{DOW}. From \refb{eqq2} we see that the separation $l$
remains finite even for $a=0$. Note, however, that when the
brane-anti-brane separation $a$ is of the order of their internal
size, the notion of distance between them becomes ambiguous, as
the presence of the brane (and the anti-brane) significantly
modifies the ambient metric. Thus $l$ is not a good measure of
the brane anti-brane separation in this range of parameters.
A more appropriate measure of the brane
anti-brane separation might be the strength of the magnetic field
that is required to hold them apart. As seen from \refb{emod5},
if we continue to use $\wt\phi$ as the azimuthal angle, then the
magnetic field $B$ does reach a finite value as $a\to 0$.
Furthermore, if we reduce $a$ further by making it negative,
keeping $\kappa$ fixed, the magnetic field continues to increase
in magnitude.
This might be taken as an evidence for further shrinkage of the
brane $-$ anti-brane separation. However, from \refb{e1}, it is
clear that in $M$-theory configurations with positive and
negative $a$ are related to each other by a simple coordinate
transformation. Thus even if we interprete the negative $a$
values as a description of smaller brane $-$ anti-brane
separation, we reach the conclusion that this configuration is
equivalent to one with
a larger brane $-$ anti-brane separation. From this point of view,
the $a=0$ solution represents
the minimum possible separation between the
brane and the anti-brane, $-$
in the same sense that the self-dual
radius represents the minimum radius of compactification of a
string theory $-$ as long as the force that holds them apart is
induced by an external magnetic field. 
Whether we call this a configuration of coincident
brane $-$ anti-brane pair or not is a matter of
convention.
Due to the fact that precisely at this point there is a choice of
coordinate system that makes the solution free from any magnetic
field, we choose to call it coincident brane $-$ anti-brane pair.

When 
the separation between the brane and the anti-brane vanishes,
the classical mass of an open string stretched between them goes
to zero in perturbative string theory. The quantum contribution 
to the mass will be of the 
order of the string scale and makes this into a tachyonic 
state\cite{BANSUS,GRGUP,PERI}. In contrast, in the $M$-theoretic
description, 
the area of the bolt, as given in \refb{e11}, reduces to a finite
non-zero answer $16\pi M^2$ for $a=0$. Noting that for $a=0$ 
the radius of
the $x^4$ direction is $R=4M$, we see that this mass is given by
$\pi R^2$.
It is instructive to express this mass in the conventional string
units. For this we note that so far in our calculation we have
set the eleven dimensional plank scale to unity, but now we can
explicitly put in that scale. In this case the mass of the
wrapped membrane is proportional to $m_p^3 R^2$, where $m_p$
denotes the eleven dimensional Plank mass. By standard
duality chasing, we can express this in string units\cite{WITD}. 
If $m_S$ denotes the string mass scale, and $g_S$ the string coupling
constant in ten dimensions, then the relevant relations are:
\be \label{e13}
m_p^3 R = m_S^2, \qquad g_S = (m_pR)^{3/2}\, .
\ee
Using these relations we see that for $a=0$ the mass of the
wrapped membrane is proportional to $m_S g_S$.

Finally we note that even though the solution does not suffer
from the instability due to tachyonic open string modes stretched
between the D-branes, there are other instabilities in the
solution. In particular, since it represents a brane $-$
anti-brane pair balanced precariously in a magnetic field, a
slight relative displacement of the pair, without an accompanying
change in the magnetic field, will throw the configuration off
balance\cite{DOW}. This, however, represents a known physical
phenomenon, and does not correspond to any new feature of the
brane $-$ anti-brane interaction.

\section{Brane on an Orientifold Plane}

An orientifold six plane of type IIA string theory is described
in $M$-theory\cite{SEIWITTH,ENH}
by the Atiyah-Hitchin metric\cite{AH,GIBMAN}:
\be \label{e14}
ds^2 = -dt^2 + \sum_{m=5}^{10}dy^mdy^m + (8m)^2 ds_{AH}^2\, ,
\ee
\be \label{e14a}
ds_{AH}^2 = 
f(\rho)^2 d\rho^2 
+ a(\rho)^2 \sigma_1^2 +
b(\rho)^2 \sigma_2^2 + c(\rho)^2 \sigma_3^2\, ,
\ee
where $f$, $a$, $b$ and $c$ are functions defined in
ref.\cite{GIBMAN}, 
\be \label{e15}
\rho=r/8m\, ,
\ee
and,
\ben \label{e16}
\sigma_1 &=& -\sin\psi d\theta + \cos\psi
\sin\theta d\phi\, , \cr
\sigma_2 &=& \cos\psi d\theta + \sin\psi
\sin\theta d\phi\, , \cr
\sigma_3 &=& d\psi + \cos\theta d\phi\, . 
\een
The coordinate $x^4$ is related to $\psi$ as:
\be \label{e17}
x^4 = 16 m \psi\, .
\ee
The coordinate ranges are given by $\pi \le \rho < \infty$,
$0\le \theta\le \pi$, $\phi$ is periodic with period $2\pi$, and
$\psi$ is periodic with period $2\pi$. Finally, there are two
identifications under the transformations $I_1$ and
$I_3$ given by:
\be \label{e18}
I_1: \quad (\rho, \theta, \phi, \psi) \to 
(\rho, \pi-\theta, \pi+\phi, -\psi)\, ,
\ee
\be \label{e19}
I_3: \quad (\rho, \theta, \phi, \psi) \to (\rho, \theta, \phi,
\psi+\pi) \, .
\ee
Our interest will be in a configuration of an orientifold plane,
together with a D- six brane on top of it. According to
ref.\cite{SEIWITTH}, the $M$-theory background describing this
configuration is simply the double cover $\wb\NN$
of the Atiyah-Hitchin space, which is obtained by modding out the
space \refb{e14a} by the transformation $I_1$, but not by $I_3$.
In order that the asymptotic radius of the $x^4$ direction
still remains equal to $16\pi m$, we need to modify 
eqs.\refb{e14}, \refb{e15} and \refb{e17} to 
\be \label{e14b}
ds^2 = -dt^2 + \sum_{m=5}^{10}dy^mdy^m + (4m)^2 ds_{AH}^2\, ,
\ee
\be \label{e20}
\rho= r/4m, \qquad x^4 = 8m\psi\, .
\ee
This description is valid in the strong coupling limit, when the
size $m$ of the manifold is large compared to the eleven
dimensional Planck scale. 

The space $\wb\NN$ admits an anti-self-dual harmonic two form
$\Omega$, given by\cite{GIBRUB,MANSCH,SEDU}:
\be \label{e20a}
\Omega = F(\rho) \Big(d\sigma_1 - {fa\over bc} d\rho\wedge
\sigma_1\Big)\, ,
\ee
where
\be \label{e20b}
F(\rho) = F_0 \exp \Big( -\int_\pi^\rho {fa\over bc}d\rho'\Big)\, .
\ee
$F_0$ is a constant.
Thus we can define a $U(1)$ gauge field on the world volume of
this system by decomposing the three form field $C$ of $M$-theory
as
\be \label{e20c}
C(t,y,x) = A(t,y) \wedge \Omega\, .
\ee
This gauge field $A$ can be identified with the $SO(2)$ gauge field
living on the world volume of the D-brane $-$ orientifold plane
system. Note that $\Omega$ is odd under $I_3$ and hence is
projected out on $\NN$. This explains why there is no gauge field
living on an isolated orientifold plane.

We now address the fate of the open string stretched 
between the D-
six brane and its image. In type IIA string theory, the classical 
contribution to this mass vanishes when the D-brane coincides with the
orientifold plane, but there is an
oscillator contribution of the order of the string scale. 
Clearly in $M$-theory this state should be
described by a membrane wrapped on an
appropriate two cycle. In this case the
most obvious choice is again the bolt described by the surface
$\rho=\pi$ in the space $\wb\NN$. The correct coordinate system
around the bolt are the new angular coordinates $\wt\theta$,
$\wt\phi$, $\wt\psi$, and the shifted radial coordinate $\wt\rho$
defined through the relations:
\ben \label{e21}
\wt\rho &=& \rho - \pi \cr
\sigma_2 &=& -\sin\wt\psi d\wt\theta + \cos\wt\psi
\sin\wt\theta d\wt\phi\, , \cr
\sigma_3 &=& \cos\wt\psi d\wt\theta + \sin\wt\psi
\sin\wt\theta d\wt\phi\, , \cr
\sigma_1 &=& d\wt\psi + \cos\wt\theta d\wt\phi\, . 
\een
In this coordinate system, the metric near the bolt takes the
form
\be \label{e22}
ds_{AH}^2 \simeq d\wt\rho^2 + 4\wt\rho^2(d\wt\psi + \cos\wt\theta
d\wt\phi)^2 + \pi^2 (d\wt\theta^2 + \sin^2\theta d\wt\phi^2)\, .
\ee
The identification under $I_1$ implies that $\wt\psi$ has period
$\pi$. Thus the metric is non-singular at $\wt\rho=0$ and
represents the product of a plane and a sphere. The area of the
sphere spanned by $\theta$ and $\phi$, after being rescaled by 
$(4m)^2$ as is required by eq.\refb{e14b}, is given by:
\be \label{e23}
A = 64 \pi^3 m^2\, .
\ee
Since the radius $R$ of the fourth direction is given by $8m$, we see
that the area is proportional to $R^2$. A membrane wrapped around
this sphere will have a mass proportional to $m_p^3 R^2$.
Transformed to the string variables, this would again correspond
to a mass of order $g_S m_S$.

In order to verify that the membrane state that we have
considered really represents the open string state that we are
looking for, note that near the bolt, $\Omega$ is given by,
\be \label{eomega}
\Omega \simeq F_0 \Big(\sin\wt\theta d\wt\phi\wedge d\wt\theta 
-{2\over \pi^2} \wt\rho d \wt\rho \wedge (d\wt\psi +\cos\wt\theta
d\wt\phi)\Big) \, .
\ee
{}From this we see that the integral of $\Omega$ on the bolt
is non-zero. Thus the wrapped membrane state is charged under the
gauge field $A$ living on the D-brane world volume. This is
precisely what is expected of an open string stretched from the
D-brane to its image.

\section{Anomalous Gravitational Coupling of the D-brane and the
Orientifold Plane}

It has been known for sometime\cite{BSV,IBR} that a D- six brane
in general background has an anomalous interaction term of the
form:
\be \label{e24}
\int C \wedge p_1\, ,
\ee
where $C$ and $p_1$ are the pull back of the three form gauge
field and the pontrjagin density from the space-time to the world
volume. In this section we shall derive this by representing the
six brane as the Kaluza-Klein monopole of $M$-theory.

In $M$-theory, there is an anomalous interaction term of the 
form\cite{VAFWIT}:
\be \label{e25}
\int C \wedge X_8(R)\, ,
\ee
where $X_8$ is an appropriate eight form constructed out of the
curvature tensor. We shall now evaluate this on an eleven
dimensional space of the form:
\be \label{e26}
(TN) \times K_7\, ,
\ee
where $(TN)$ denotes the Euclidean self-dual
Taub-NUT space and $K_7$ is a seven
dimensional space with Minkowski signature. Using the
identification of the Euclidean Taub-NUT space with the
transverse space of the six-brane, we can interpret \refb{e26}
as a six brane with world volume wrapped on $K_7$. Since $X_8$
contains a term proportional to $p_1\wedge p_1$, and since 
\be \label{enn}
\int_{TN}p_1=-2\, ,
\ee
\refb{e25}, evaluated
in such a background gives a term of the form:
\be \label{e27}
a \int_{K_7} C\wedge p_1\, ,
\ee
where $a$ is a constant. 

This is precisely the term that we wanted. It remains to verify
that the coefficient $a$ is given correctly, but before we do
that, let us note that a similar coupling between the 3-form field
$C$ and the curved orientifold six plane\cite{DJM} can be derived by
representing the transverse space of the orientifold plane by the
Atiyah-Hitchin space. If $L_7$ denotes the world-volume of the 
orientifold plane,  we shall get a coupling of the form:
\be \label{e27a}
b \int_{L_7} C\wedge p_1\, ,
\ee
where the constant $b$ is proportional to the integral of $p_1$
on the Atiyah-Hitchin space. 

We shall now show that this procedure yields the correct values
of the constants $a$ and $b$. To do this, let us first note that
since a D- six brane on top of an orientifold plane is
represented by the double cover of the Atiyah-Hitchin space, we
have the result:
\be \label{e28}
\int_{TN} p_1 + \int_{AH} p_1 = 2\int_{AH} p_1\, .
\ee
Thus the integral of the pontrjagin index over the Taub-NUT space
and the Atiyah-Hitchin space gives the same answer. This gives
$a=b$, as was found in ref.\cite{DJM} for the orientifold six
plane. 

The overall normalization is fixed by noting that $\int_{TN}p_1$
given in \refb{enn}
is (1/24) times $\int_{K3} p_1$. 
Thus $a$ and
$b$ are (1/24) times the coefficient of the $\int C\wedge p_1$
term for $M$-theory on K3.
To see that this is the correct answer, consider type IIA of 
$T^3/(-1)^{F_L}\Omega \II_3$, where $\II_3$ denotes the reversal
of sign of all the three coordinates on the torus, $(-1)^{F_L}$
changes the sign of Ramond sector states on the left, and
$\Omega$ is the world-sheet parity transformation. This theory
has 16 D-branes and 8 orientifold planes, so the net contribution
to the $C\wedge p_1$ term has a coefficient equal to $24a$,
{\it \i.e.} equal to that of K3. But since this theory is dual to
$M$-theory on K3, this is precisely the correct answer.

{\bf Acknowledgement}: I wish to thank D. Jatkar and S. Mukhi for
useful discussions.

\end{document}